# Muon ID at the ILC

C. Milsténe, G. Fisk, A. Para
*FNAL, Batavia, IL 60510, USA*

This paper describes a new way to reconstruct and identify muons with high efficiency and high pion rejection. Since muons at the ILC are often produced with or in jets, for many of the physics channels of interest[1], an efficient algorithm to deal with the identification and separation of particles within jets is important. The algorithm at the core of the method accounts for the effects of the magnetic field and for the loss of energy by charged particles due to ionization in the detector. We have chosen to develop the analysis within the setup of one of the Linear Collider Concept Detectors adopted by the US. Within b-pair production jets, particles cover a wide range in momenta; however ~ 80% of the particles have a momentum below 30 GeV[2]. Our study, focused on bbar-b jets, is preceded by a careful analysis of single energy particles between 2 and 50 GeV. As medium energy particles are a substantial component of the jets, many of the particles lose part of their energy in the calorimeters and the solenoid coil before reaching the muon detector where they may have energy below 2 GeV. To deal with this problem we have implemented a Runge-Kutta correction of the calculated trajectory to better handle these lower energy particles. The multiple scattering and other stochastic processes, more important at lower energy, is addressed by a Kalman-filter integrated into the reconstruction algorithm. The algorithm provides a unique and powerful separation of muons from pions. The 5 Tesla magnetic field from a solenoid surrounds the hadron calorimeter and allows the reconstruction and precision momentum measurement down to a few GeV.

## 1. INTRODUCTION

We describe a muon identification procedure that involves a comparison of the charged track trajectory reconstructed in the tracking systems with the location of energy deposits detected in the calorimeters and the muon detectors. This comparison requires that the track trajectory is extrapolated beyond the tracking detectors. Our initial extrapolation method, the *swimmer*, was based on a helical extrapolation of the track assuming constant angular velocity. While this is an acceptable approximation for high momentum tracks, it fails at the low and medium momentum end of the spectrum where the energy loss and multiple scattering cause significant deviation of the actual track trajectory from an ideal helix, the discrepancy being amplified by the bending effect in the strong magnetic field.

The *stepper*, an improved software package, uses an analytical procedure to propagate the charged track determined from the vertex and silicon tracking systems through the calorimeters and solenoid in small steps accounting for dE/dx and $q\vec{v} \times \vec{B}$ effects [3] at each step. The *stepper* improves the identification and reconstruction efficiency of low energy muons from 0.06% at 3GeV to 33% (see table 2). When a Runge-Kutta correction is applied to the *stepper* algorithm the match between the projected track and hits in the calorimeters and muon detector further improves the identification efficiency to 66%.

Finally a Kalman filter *stepper* has been developed which improves further the results by taking into account multiple scattering, Bremsstrahlung and decays. This paper describes the general principles of the muon reconstruction package and its implementation.

The International Linear Collider (ILC) is a proposed electron-positron accelerator complex to facilitate the study of electroweak symmetry breaking at a center-of-mass energy of 500 GeV. The e+ and e- beams are separately accelerated to high energy by a series of ~ 4000 superconducting (SC) RF cavities in 15 km tunnels on either side of the interaction point (IP) where they are then brought into collision via a beam delivery system.

The IP is surrounded by a collection of detectors that track and measure the energies of the out-going particles created in the e+e- collisions. The detectors are arranged to accurately measure the position and direction of outgoing charged particles, their momenta, and their energies and to identify hadrons, leptons and missing energy that is typically the hallmark of neutrinos that interact so weakly that they leave none of their energy in the detectors. Electromagnetic and hadronic calorimeters are used to identify electrons/photons and hadrons, respectively. Muons are identified by their penetration through the tracking and calorimeter detectors where they lose small amounts of detectable energy by their ionization of the material media through which they pass. The muon identification proceeds primarily via the observation of ionization energy that is observed after electrons, photons and hadrons have converted all their energy in upstream material.

We will start by a short description of the detector which will be followed by a detailed study of the algorithm and its improvements described above.

## 2. THE DETECTOR

The detector has five main sub-detectors: The vertex sub-detector is composed of a central barrel system with five layers and forward systems composed of four disks. The tracker is composed of five cylindrical barrels and five disk-shaped end planes. The electro-magnetic calorimeters have 30 layers of 0.25cm Tungsten and 0.05cm Si-detectors (supplemented by 0.1cm air and 0.1cm G10) for a 0.5cm layer thickness; the hadron calorimeter has 34 layers each with 2cm steel and 1cm polystyrene, and finally the postulated muon detector has 32 layers with 5cm Fe plates and 32 instrumented gaps of scintillators. A total of 240cm of iron is needed for the SC magnet flux return, of which 32 layers are instrumented as a muon detector. The detector is represented in Table 1 in terms of radiation lengths, interaction lengths and dE/dx

| ECal (30 Layers) | HCal (34 Layers) | Coil | Mudet (32 Layers) |
|---|---|---|---|
| $X_0$ - 21.75 | $X_0$ - 39.44 | $X_0$ - 13 | $X_0$ - 79.5 |
| $\lambda$ - 0.872 | $\lambda$ - 4.08 | $\lambda$ - 2 | $\lambda$ - 8.4 |
| dE/dx - 190 MeV | dE/dx - 800 MeV | dE/dx - 362 MeV | dE/dx - 1400 MeV |
| <u>Segmentation</u> $\Delta\Phi = \Delta\theta = 3.7$ mr | <u>Segmentation</u> $\Delta\Phi = \Delta\theta = 5.23$ mr | | <u>Segmentation</u> $\Delta\Phi = \Delta\theta = 21$ mr |

Table1: The Number of Interaction/Radiation Lengths spanned by the Subdetectors

In figure1 is represented a transverse cut of a quarter of the detector with the different subdetectors and the number of interaction lengths they cover. A 5 GeV muon is also shown as well as the reference scale for 100cm .

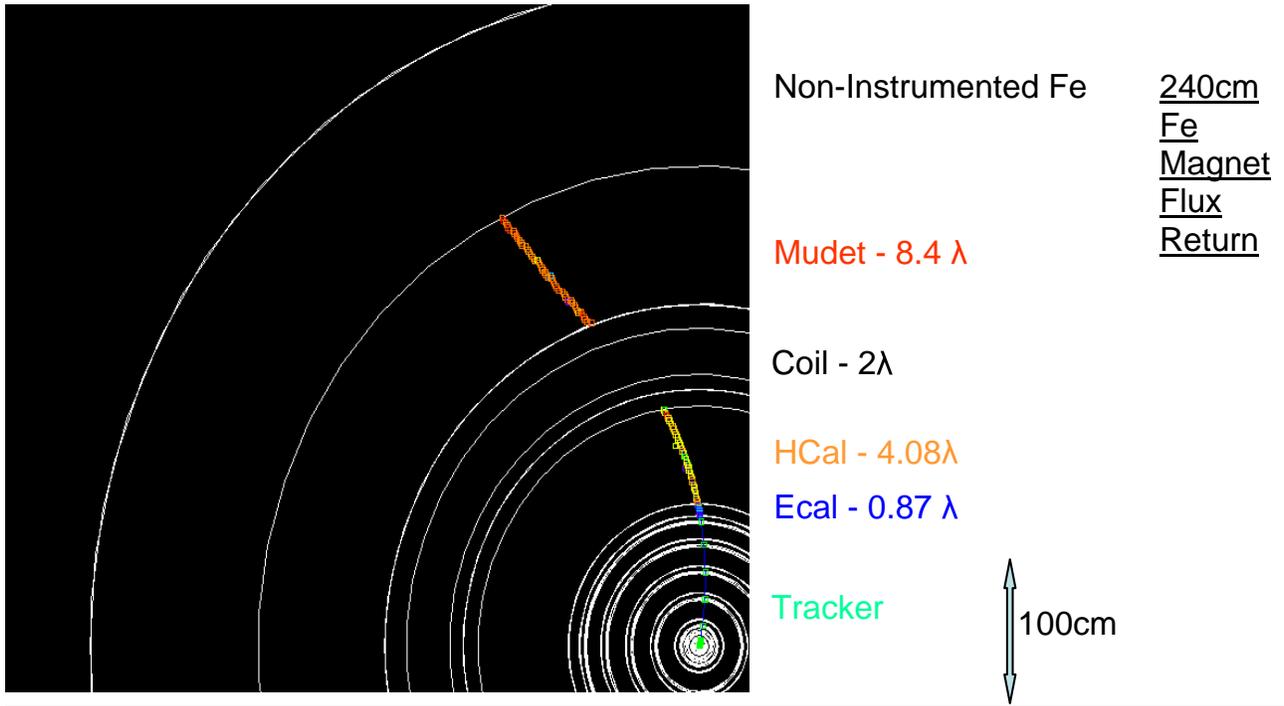

Figure 1- A transverse Cut of the Various Subdetectors with a 5 GeV Muon Trajectory Recorded.

## 3. THE STEPPER

The stepper starts with a particle at the IP and it computes step-wise the particle trajectory throughout the complete detector. A uniform axial magnetic field $B_z$ is assumed up to the coil itself. The momentum components $p_x$ and $p_y$ undergo changes due to the q $\mathbf{v} \times \mathbf{B}$ term whereas the energy loss in material contributes to a reduction of all the components of particle momentum.

### 3.1. The Parametrization

Each component of the momentum changes at each step. There is a momentum change $\delta p_x(B_z)$, $\delta p_y(B_z)$, due to the magnetic field, and one of $\delta p_x(material)=\gamma_x$, $\delta p_y(material)=\gamma_y$, $\delta p_z(material)=\gamma_z$ due to energy loss by ionization in material.

In the equations below, q is the charge, $B_z$ the magnetic field, dt(n) the time spent and ds the path length in one step. The algorithm for determining the particles trajectory including energy loss is given by the equations below. In these equations, q is the charge, $B_z$ the magnetic field, dt(n) the time spent and ds the path length in one step.

$$p_x(n+1) = p_x(n) - 0.3 * q * \frac{p_y}{E(n)} * c_{light} * B_z * \delta t(n) - \gamma_x(n) \;;$$

$$p_y(n+1) = p_y(n) + 0.3 * q * \frac{p_x}{E(n)} * c_{light} * B_z * \delta t(n) - \gamma_y(n) \;;$$

$$p_z(n+1) = p_z(n) - \gamma_z(n) \;;$$

$$\gamma_i(n) = \frac{dE}{d_i} * \frac{E(n)}{|p(n)|} * \frac{p_i(n)}{|p(n)|} * \delta s \;;\; i = x, y, z \;.$$

Mixed units are used, $p_x$, $p_y$, $p_z$ are in GeV/c, E(n) in GeV, $c_{light}$ =3x10$^8$ m/s, δt in seconds, $B_z$ in Tesla

The point ( x(n+1),y(n+1),z(n+1) ) is the position at step n+1, after the momentum change to $p_{x,y,z}$(n+1) at step n.

$$x(n+1) = x(n) + \frac{p_x(n+1)}{E(n+1)} * c_{light} * \delta t(n) \;;$$

$$y(n+1) = y(n) + \frac{p_y(n+1)}{E(n+1)} * c_{light} * \delta t(n) \;;$$

$$z(n+1) = z(n) + \frac{p_z(n+1)}{E(n+1)} * c_{light} * \delta t(n) \;.$$

In the next figure are represented the 4 stages of the particle at each step in the stepper. Those 4 stages are repeated N times if the stepper is called N times for a particular layer of the subdetector. The particle is represented by a point in the phase-space {x, y, z, $p_x$, $p_y$, $p_z$}. The 4 stages are described below

## Stepper Processing Flow

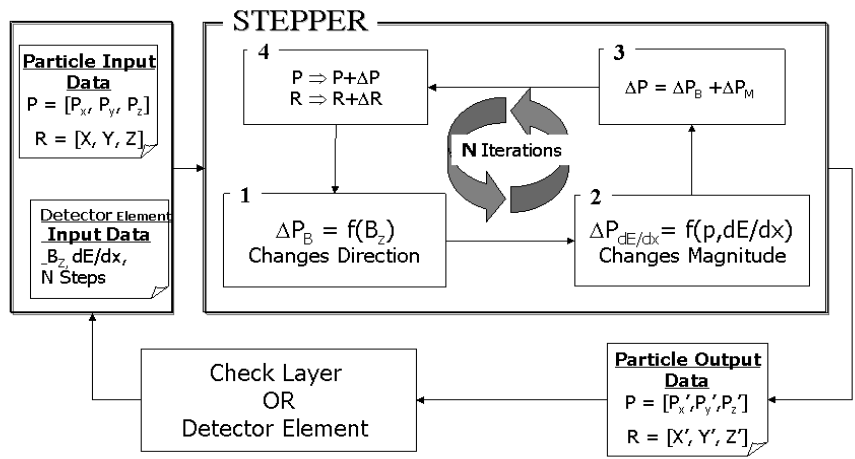

Figure 2- Stepper processing flow in Ecal, HCal and Mudet

1) $\Delta P_B$ - changes direction due to the field Bz
2) $\Delta P_{dE/dx}$ – changes magnitude due to dE/dx
3) $\Delta P = \Delta P_B + \Delta P_{dE/dx}$ (x, y, z components)
4) $P = P + \Delta P$; $R = R + \Delta R$ (x, y, z components)

The output {x, y, z, $p_x$, $p_y$, $p_z$} becomes the new input data for the next layer or the next subdetector.

### 3.2. The Muon Candidate

The muon identification algorithm requires a well fitted charged track consistent with the observed energy deposits in EM and HAD calorimeters and at least 12 hits /12 layers within an extrapolated ($\theta$, $\varphi$) road in the muon detector. Calorimetric cells are associated with a track if they are within 2 angular bins in the EM calorimeter and the muon detector or 3 angular bins in the HAD calorimeter, ( $\Delta\varphi = \Delta\theta = \pi/840$ in ECal , $\Delta\varphi = \Delta\theta = \pi/600$ in HCal, $\Delta\varphi = \Delta\theta = \pi/150$ in the muon detector). The current version of the muon ID package is restricted to the barrel detector, defined as $0.95\,\text{rad} < \theta < 2.2\,\text{rad}$.

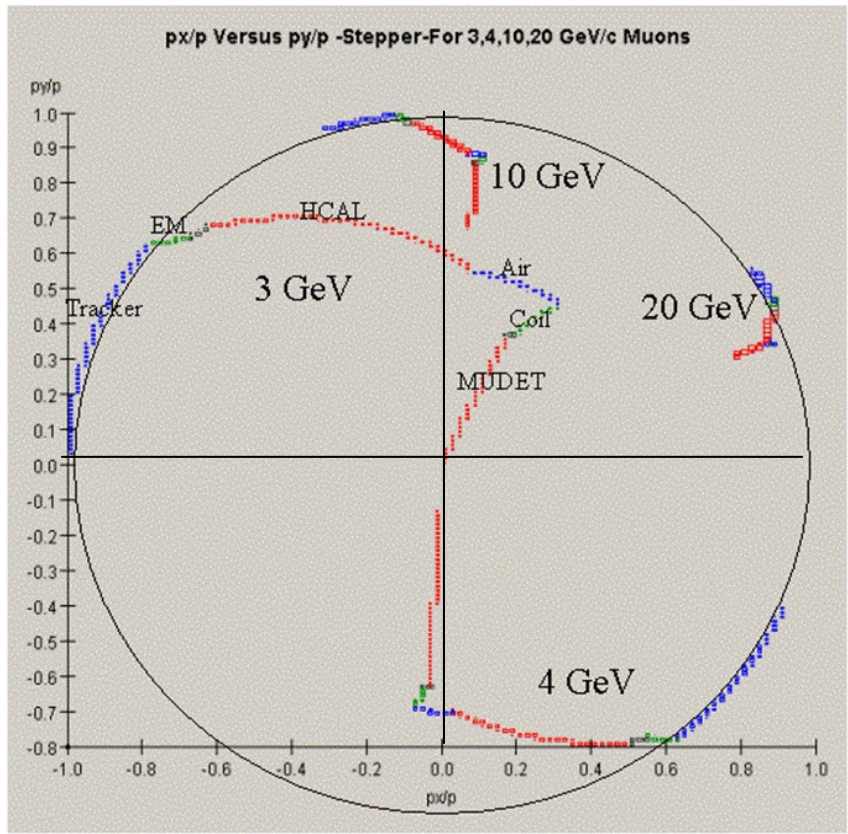

Figure 3- Evolution of ($p_x/p_{IP}$, $p_y/p_{IP}$) through the sub-detectors, with $p_{IP}$, the magnitude of the particle momentum at the IP.

Figure 3 shows muons of different energies produced perpendicular to the beamline (neglecting the small crossing angle of the beams). In this case the particle trajectory takes place in the transverse plane. It is shown here in the ($p_x$, $p_y$) plane to convey the essential information. In the vertex detector and the tracker, where the amount of material is negligible, we see that $\sqrt{(p_x * p_x + p_y * p_y)} / p_{IP} \sim 1$. It is the direction of the momentum which varies (as a result of the magnetic field) and not its magnitude; therefore, the track remains on the circumference of the circle. As the particle enters the calorimeters the amount of material encountered increases significantly and the particle starts to lose energy. Correspondingly, the trajectory in $(p_x/p_{IP}, p_y/p_{IP})$ plane moves toward the center of the circle the radius decreases. The radius represents the ratio between the magnitude of the particle momentum to its magnitude at the IP.

As the particle enters the HCal it loses even more energy than in Ecal since ($\Sigma((dE/dx)*\Delta L)$) is bigger and it goes on losing energy in the coil and Mudet.

In Mudet, because the magnetic field is inverted and smaller in magnitude, the particle changes its bending direction. For energies at or below 3 GeV/c., the point $(p_x/p_{IP}, p_y/p_{IP})$ ends up at the center of the circle the particle having lost all its energy by dE/dx. The 4 GeV/c muon is left with ~10% of its momentum at the limit of the instrumented muon detector, and the higher the muon momentum the smaller the proportion of momentum loss, as can be seen in the plot for 10GeV/c and 20 GeV/c Muons. For a particle coming in with a shallow angle, e.g. $\theta \sim 0.95$ rad the path-length, $\Delta L$, is increased by almost a factor of 2, ($1/\cos\theta$) and so is the loss of energy by ionization $((dE/dx)*\Delta L)$. This affects particles in the range from 3 to 5 GeV which end up with a momentum close to zero depending on the amount of material they pass through.

### 3.3. Improvement in Track Reconstruction

We have compared the performance of the muon identification algorithm using the *stepper* with that of the original package of R. Markeloff which was using the standard *swimmer* for the track extrapolation to the calorimeters and muon detectors. The *swimmer* approach does not account for the energy loss in the calorimeters, hence it fails for low and medium momenta muons. We have improved its performance for the low momentum tracks by introducing momentum-dependent cuts, $(\delta\varphi, \delta\theta \sim 1/p)$, on the distance between the extrapolated track and the calorimeter/muon detector cell. This version of the package is denoted as 'Swimmer + Ad Hoc dE/dx' in the figure below..

In Figure 4 and Table 2 are shown the muon–ID efficiency improvement with the evolution of the algorithm for single muons from 3GeV to 50 GeV.

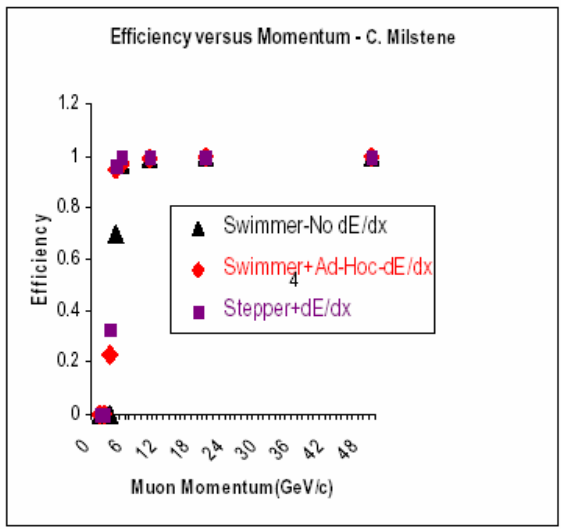

| E(GeV) \ Techn. | 3 | 4 | 5 | 10 |
|---|---|---|---|---|
| No dE/dx | 0.06% | 70% | 97% | 99.% |
| Ad-Hoc dE/dx | 23% | 95% | 97% | 99.% |
| V x B + dE/dx | 33% | 96% | 99% | 100% |

Figure 4 & Table 2: Muon-ID Efficiency as a function of the momentum

The histogram in Figure 5 shows the angular deviation of the observed hit from the extrapolated track trajectory in φ at different depths (layer number) in the muon detector. Δ(φTrack-φHits) for 20 GeV muons (left) is typically ~one φbin wide whereas Δ(φTrack-φHits) for a 4 GeV muon is ~ 4 times larger. The remaining inefficiency at low muon momentum is primarily due to the stochastic nature of processes not accounted for by the stepper, e.g. multiple scattering, Beamstrahlung or decays.

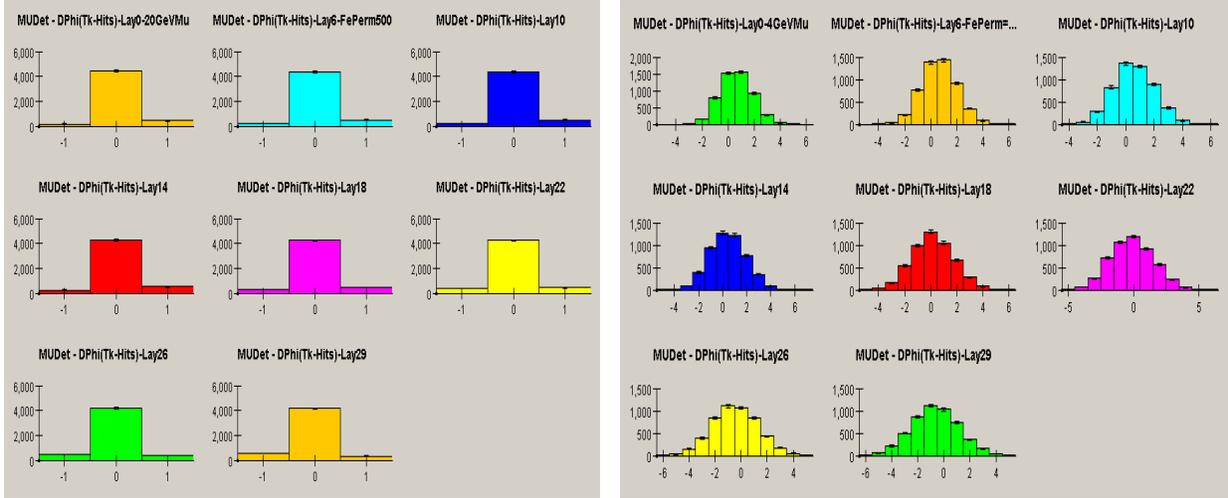

Figure 5: Angular Resolution in φ in MuDet at different depths; left: 20 GeV Muons, right 4 GeV Muons.

## 4. THE MUON IDENTIFICATION IN JETS

Reconstruction and identification of single muons is a relatively easy task. The more difficult, albeit more realistic, case involves muon identification in high energy jets. In such an environment, in addition to the issue of the identification efficiency, we encounter an issue of the purity of the 'muon' sample. Most of particles in jets are hadrons, hence even with small probability of misidentification the fake muons may overwhelm the muon candidates . To study these issues we have generated a sample of 10,000 b-quark pair events produced in $e^+e^-$ collisions at the center of mass energy up to 500 GeV.

Table 3 shows a breakdown of generated charged particles in this sample. Particles below 3 GeV do not penetrate the muon detector, hence they cannot be identified as muons by our algorithm.

We also notice that only 34% of the pions in the table have a momentum at or above 3 GeV/c whereas almost 70% of the muons are produced at or above 3 GeV/c in heavy flavor b-jets and will therefore reach the muon detector.

In absolute value though, the number of pions above 3 GeV which can reach the muon detector overcomes the number of muons by a factor 16 and the number of kaons by a factor 4. Of course the fact that hadrons interact in HCal will be helpful in filtering them out from Mudet.

In figure 6 is shown the momentum distribution of muons and pions generated at or above 3 GeV/c in bbar-b jets and those pions and muons after detection using our algorithm. One can see that among those 787 muons above 3GeV/c, almost 300 have a momentum below 7.5 GeV/c, of which 200 muons have a momentum even below 5 GeV. Most of those muons, depending on their incoming azimuthal angle will be left with less than 2 GeV/c at the end of their trajectory in the muon detector.

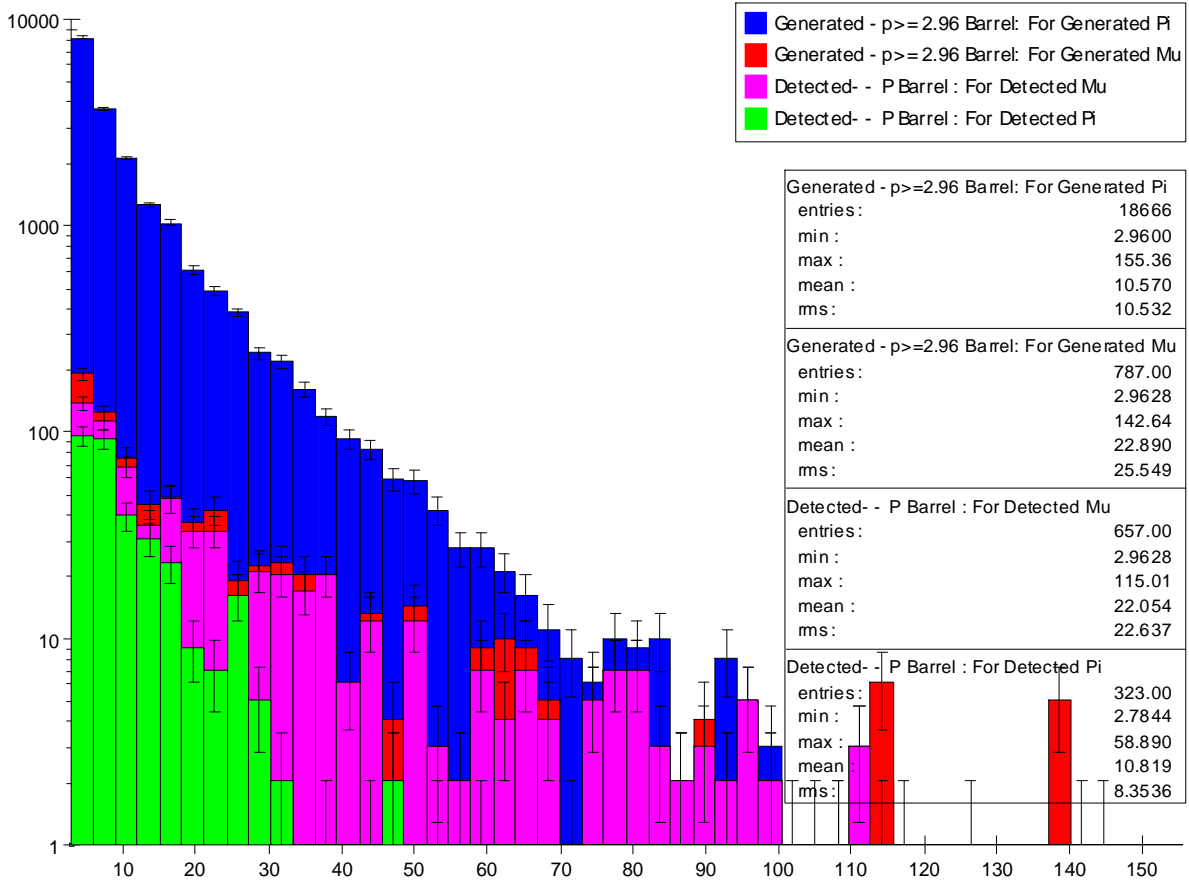

Figure 6- Generated/Detected Muons and Pions Momentum (GeV) Distribution in 10,000 bbar-b

|  | π | K | Protons | Muons |
|---|---|---|---|---|
| Total Generated | 55805* | 8310* | 2816* | 1147 |
| Generated .>3GeV | 18666 | 4473 | 1622 | 787 |
| Fraction >3GeV | 34% | 54% | 58% | 70% |
| Reconstructed Tracks > 3GeV | 18024 | 4304 | 1614 | 739 |

Table 3: Muons and Hadrons breakdown of Generated particles within bbar-b

## 4.1. Runge-Kutta Correction

We now focus mainly on the simulated muons of lower momenta, which are ~30% of the data that reach the muon detector (3GeV≤ p <5GeV) and for which the approximation $\Delta p_T / \Delta t \sim dp_T / dt$ is insufficient, at least at the end of their path. Here $\Delta p_T$ (GeV/c) is the variation in the transverse momentum of a particle going through a magnetic field B (Tesla) for a time $\Delta t$(s) and

$$dp_T / dt = \alpha \bar{v} \times \bar{B} , \text{ with } \alpha = 0.3q(1/E)c_{light}dt$$

E is the particle energy in GeV, q its charge in electron units and $c_{light}$ (m/s). In a 5 Tesla magnetic field, for low momenta, one has to calculate the integral in order to obtain the finite difference equation of motion:

$$\int \alpha \bar{v} \times \bar{B} \, dt ,$$

and the magnetic field dependent change in momentum is now given by the equations

$$\Delta p_x B = (\alpha / \delta).p_y.B_z - \eta.p_x$$
$$\Delta p_y B = (\alpha / \delta).p_x.B_z - \eta.p_y$$
$$\Delta p_z B = 0.$$

This introduces correction terms in the magnetic field dependent momentum change.

$$\delta = 1 + 0.25\alpha^2 B^2 , \eta = 0.5\alpha^2 B^2$$

The term - $\eta \cdot p_i$ ; (i=x ,y) is a friction term. The energy loss due to ionization in matter (for step n of the trajectory) is given by $\Delta p_{i\,Matter} = (dE(n)/dx).(E(n)/p).(p_i(n)/p(n)).ds$, $i = x,y,z$

Here *ds* is the distance traversed by the particle.

## 4.2. Muon Identification with the Improved Method

A significant fraction of hadron-induced showers produce charged particles that reach the muon detectors where they are classified as muons by the muon identification algorithm described earlier. In figure 6 one observes that 323 pions passed the algorithm and are identified wrongly as muons (in green on the figure). The number of such false muon candidates has increased with improved reconstruction. More of the medium and low energy particles get reconstructed, which further improves the detection efficiency that has already increased by using the *stepper*. This rate can be reduced by exploiting the patterns of hadron showers in HCal compared to the minimum ionizing signals left by muons.

The muon has two main characteristic properties:

1) The muon leaves a repetitive pattern (1 to 2 hits per cell) in all sub-detectors.
2) The muon travels through the sub-detectors without strong interactions.

The hadrons do not typically exhibit these properties. Hadrons characteristically interact, which gives them an irregular hit pattern during part of their trajectory in HCal, and as a consequence, they don't reach the end of the hadron calorimeter. To improve our cuts, we eliminate those hadrons that interact in HCal. To eliminate those hadrons that do not interact in HCal we use the muon detector as discussed below.

### 4.2.1. The Extended Algorithm

The extended analysis starts by requiring as before, reconstructed and well fitted tracks from the tracker with energy deposited in ECal, HCal and Mudet within a (θ,φ) angular road around their path. The angular width is optimized for each sub-detector [3] [4] . For tracks below 10 GeV a (1/p) dependent bin width has been chosen.

We reconstruct the track with the **vXB** and dE/dx effects for each particle using the Runge-Kutta improved *stepper*. We require no more than 2 hits per cell and that the particle reaches a given depth (12 layers) in the muon detector.

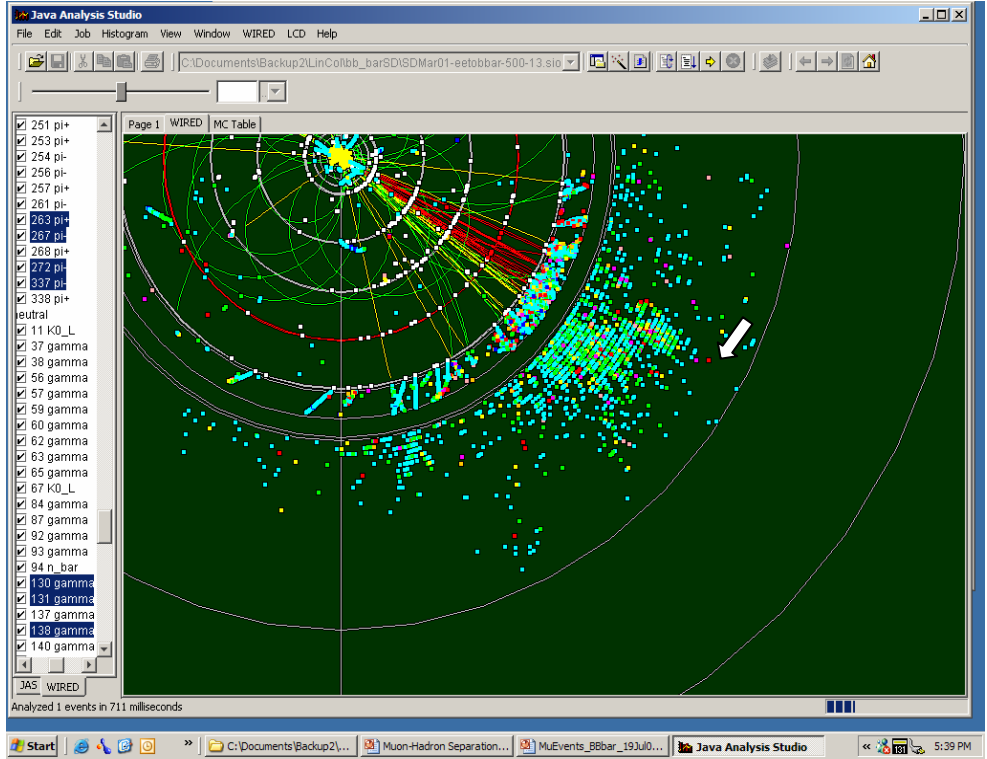

Figure 7: Hadronic shower in a jet projected on the detector transverse (x, y) plane (cm)

We exploit the fact that the hadronic showers deposit most of their energy in the innermost layers of the hadron calorimeter. This is illustrated in Figure 7 where among the red track are pions of 32 and 37 GeV which interact in HCal so that there are no dE/dx deposits in the last layers of HCal as indicated by the white arrow in Fig. 7. Projected tracks that do not leave a signal in the last 5 layers of HCal are rejected as muons. However, we also require not more than 4 hits/layer to reject charged particles that result from interactions of nearby neutral hadrons (neutrons and $K_0$'s). Therefore, the muons that reach the end of HCal must leave some hits in the 5 last layers. In this scan along the projected track we reject muon candidates where there are no hits (i.e. no dE/dx losses recorded) in 2-3 consecutive layers of the HCal or the muon detector. An interacting hadron displays a set of hits that typically ends in a splash followed by no hits. These extra requirements do not lead to any losses of genuine muons, but they reduce the number of hadrons that would be identified as muons without such requirements. In the muon detector we request that the minimum of hits/layer is smaller or equal 2, since interacting hadrons might have three and more hits in each layer of Mudet, and the maximum of hits per layer to be below 7, this allow not to cut muons which have neutral particles in their vicinity which come from neighbor hadrons.

### 4.2.2. Performance of Muon ID

The effect of the cuts is shown in Table 4. The detection efficiency is defined as the ratio of the total number of muons detected by the total number of reconstructed muons projected from the tracker where they have passed a 'good fit' criteria. The detection efficiency defined this way under-estimates true efficiency because, due to multiple scattering, tracks can leave the fiducial volume of the barrel detectors and pass into the forward detector regions where muon tracking identification has not yet been developed. There are also cases for which the track is not detected out to the 12$^{th}$ layer because it reached the end of its range before the required 12th layer. This is true for many medium energy shallow tracks. It should also be noted that 12 pions and 3 K's that decay to muons are properly identified by our algorithm as muons in the muon detector.

| Detector | Muons | Pions | Kaons | Protons |
|---|---|---|---|---|
| Tracker Recons. Final | 739 | 18024 | 4303 | 1614 |
| Tracker Good Fit | 715 | 17120 | 4072 | 1579 |
| HDCal $1 \leq$ min Hits $\leq 2$ | 700 | 588 | 247 | 26 |
| 5 –Last Layers > 0 Hit | 700 | 357 | 204 | 15 |
| MUDet, $\geq 12$ hits $\geq 12$ layers | 671 | 77 | 50 | 5 |
| MUDet Min Hits $\leq 2$ ; Max Hits $\leq 7$ | 670 | 59 | 39 | 5 |
| Efficiencies of $\mu$ Detection And Hadrons Rejection | Detect. 94% | Rej. 1/305 | Rej. 1/110 | Rej. 1/342 |

Table 4: Muon Identification in 10000 b-bbar with at least 12 layers/12 Hits in the Muon

The overall muon detection efficiency obtained without discounting for the effects described above is ~94%. This covers the entire momentum range. The efficiency is ~100% at, and above, 10 GeV. For muon momentum at 3GeV/c the detection efficiency of 0.06% with the helical *swimmer* becomes 33% with our basic *stepper* algorithm, and is 66% with the Runge-Kutta corrected *stepper*. The total number of muons reconstructed increases from 657 before the Runge-Kutta correction to 670 muons after, mostly in the medium and lower energy range. At the same time the number of pions misidentified as muons moves from 323 (figure 6) to 59 (table 4). When a Kalman filter is used to account for multiple scattering and other random processes we find a better separation of muons from hadrons.

## 5. THE KALMAN FILTER IMPLEMENTATION OF THE STEPPER

We expect that further improvement of the muon identification efficiency can be accomplished if the predicted trajectory of the muon can account for actual occurrences of stochastic processes, like multiple scattering. At the same time we expect that background rejection capabilities can be improved if the decay kinks can be detected, although this may not be possible.

Both of these goals can be achieved, in as much as it is possible, by the application of a Kalman filter technique. Its main advantage over the other methods of track extrapolation stems from its ability to use actual measurements to adjust local track parameters and to offer a better prediction for the track trajectory.

### 5.1. The Principle

The Kalman Filter is a method used to construct processes for which the information at each step can be fully derived from the information at the preceding step and where a covariant matrix of error is available at each step. The Kalman Filter is composed of two components. The propagation component, using the propagation matrix, propagates the information and the error information at each step. The filtering component combines the information received at the end of the step to the information from data measurements.

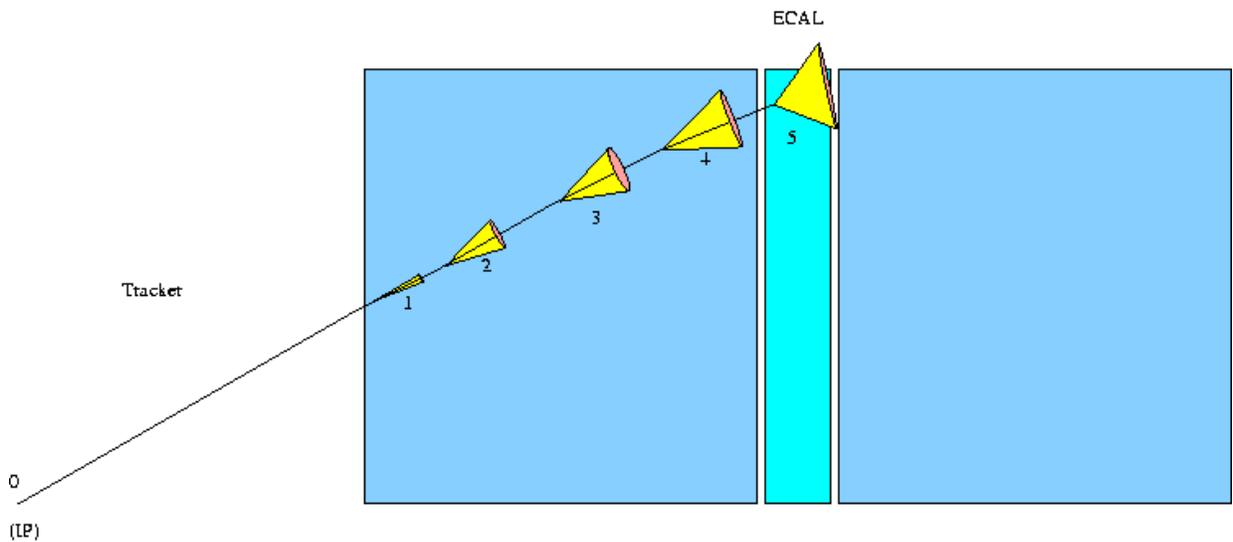

Figure 8: Stepping in the Passive Material (1, 2, 3,4, ) and recording in the Active Material step 5

The Kalman Filter operates by using a "state vector" that describes the processes' history. For the present case, a phase space point $\{x, y, z, p_x, p_y, p_z\}$, is used as the state vector, The state vector change includes the effects of dE/dx and of the magnetic field in a step by step procedure. It is calculated using the transport matrix, steps 1, 2, 3, 4 shown in Figure 8, in the passive material. The same analytical form is used in the stepper, but is now translated into a transport matrix, with the Runge-Kutta term included. The state vector at location k-1 is propagated using the propagation matrix, together with the covariant matrix, represented by the cone, which provides the information on random processes,. Following the propagation step, the state vector is updated using Kalman filtering. The filtering combines the information derived from the propagation with the measurement made at the new point to produce an optimal state vector. It takes place when the hit is recorded in the <u>active material</u> (step 5) as shown in Figure 8. The procedure defines, for each radius, a $(\varphi, \theta)$ dynamic path in which the hits left by the particle in the active material of the sub-detectors are the data collected. The number of steps was optimized for the *stepper* at 10 steps per layer by considering

**12**

both the reconstruction efficiency and running time for all the subdetectors. Accounting for the fact that we get combined information on the phase-space point and the covariant matrix at each point, we have obtained good results with half the number of steps.

## 5.2. The Equations

The stages of the Kalman Filter are described by the equations below.

$$\begin{cases} \vec{x}_k(-) = \Phi_{k-1} \cdot \vec{x}_{k-1}(-) \\ P_k(-) = \Phi_{k-1} \cdot P_{k-1}(-) \cdot \Phi_{k-1}^T + Q_{k-1} \end{cases}$$ (1) Propagation in passive materiel

$$\begin{cases} \vec{x}_k(+) = \vec{x}_k(-) + K_k \cdot [\vec{z}_k - H_k \cdot \vec{x}_k(-)] \\ P_k(+) = [1 - K_k \cdot H_k] \cdot P_k(-) \end{cases}$$ (2) K.Filter, in scintillator,

where the signal is recorded

$$K_k = P_k(-) \cdot H_k^T \cdot [H_k \cdot P_k(-) \cdot H_k^T + R_k]^{-1}$$ (3) K.Gain Matrix

$$Q_{k-1} = |\vec{p}| \cdot \Theta_0 \cdot I$$

$Q_k$ is the noise from Multiple scattering - a [6x6] matrix, $R_k$ is the measurement error, dx, dy, dz - a [3x3] matrix $H_k$ is the measurement matrix - a [3x6] matrix, $\Phi_k$ = propagation matrix, applied in passive material- [6x6] The matrix, $X^k(-)$ is the extrapolated vector state (x, y, z, $p_x$, $p_y$, $p_z$) -[6 x 1], $Z_k$ is the measured quantities ($\Phi,\theta,r$) translated into (x, y, z)-[3 x 1] matrix, $X_k(+)$ is the state vector after applying the Kalman filter-[6x1] Matrix, $K_k$ is the Kalman Gain matrix - a [6 x 3] Matrix.

$$\begin{Bmatrix} x_k(-) \\ y_k(-) \\ z_k(-) \\ px_k(-) \\ py_k(-) \\ pz_k(-) \end{Bmatrix} = \left( dT * \begin{pmatrix} aa & ab \\ ba & bb \end{pmatrix} + I \right) * \begin{Bmatrix} x_{k-1}(+/-) \\ y_{k-1}(+/-) \\ z_{k-1}(+/-) \\ px_{k-1}(+/-) \\ py_{k-1}(+/-) \\ pz_{k-1}(+/-) \end{Bmatrix} ; \quad \phi = dT * \begin{pmatrix} aa & ab \\ ba & bb \end{pmatrix} + I$$

$$ab = \begin{pmatrix} (cdedx) & f(Bz) & 0 \\ -f(Bz) & (cdedx) & 0 \\ 0 & 0 & (cdedx) \end{pmatrix} ; \quad bb = \begin{pmatrix} dxyz & 0 & 0 \\ 0 & dxyz & 0 \\ 0 & 0 & dxyz \end{pmatrix} \quad ; \text{cdedx} = \text{dE/dx} \cdot 100 * \text{clight/Pabs}$$

$$R_0 = \begin{pmatrix} dx & 0 & 0 \\ 0 & dy & 0 \\ 0 & 0 & dz \end{pmatrix} ; \Theta_0 = (13.6 MeV / P \cdot \beta c)\sqrt{x/X0} \cdot (1+0.038 \cdot lan(x/X0)); \quad \beta = c = 1$$

$x = r \cdot \sin \Phi; y = r \cdot \cos \Phi; z = r \cdot ctg\Theta$

$dxyz = 100 * clight / E_n; \quad f(Bz) = 0.3 * q * \frac{p_x(n)}{E(n)} * clight * B_z \quad dis \tan ce \ in \ (cm), \ energy \ in \ GeV, \ B(Tesla)$

The transport matrix is given below in some detail. The matrix represents the propagation equations given in [3] and [4], dT is the time taken by step n., I is a [6x6] matrix Unity. The term - $\eta \cdot pi$ ; i=x ,y is a friction term is in the diagonal in the Transport Matrix. With $\theta_0$ the multiple scattering [5].

The state vector at location k-1, is propagated using the propagation matrix, to location k. The choice of the state vector as the phase space point allows the use the stepper Algorithm translated into the propagation matrix. In the notation adopted the (-) indicates the extrapolation and the (+) the Kalman weighting.

### 5.3. The Results

The results for single muons at 4GeV before and after the application of the Kalman Filter are shown in Figure 9.

The figure below shows Δφ (track-hit) between the extrapolated track using the stepper and the hit at different depths (different layer) in the Muon Detector.

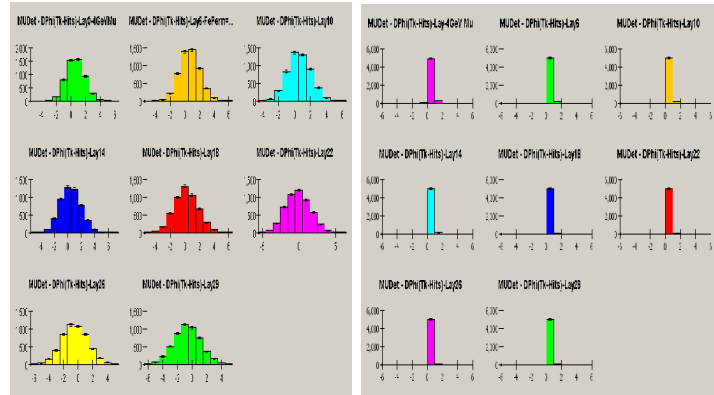

Figure9: Δ(φtrack-φhit) at 4 GeV Before and After applying Kalman

In the LHS of the figure the width of the distribution is driven by the resolution at 4 GeV and the hit collection path covers ~ 3bins in φ. The RHS of the figure shows that one can restrict the hit collection to one bin in φ. The filter helps to master random effects such as multiple-scattering when particles lose sufficient momentum and enter the domain where these effects are dominant. As a result, the filter allows a better separation between muons and neighboring hadrons in jets since the effect of multiple-scattering and the kinks are corrected each layer at a time and this in turn corrects for the distances between the reconstructed track and the hits.

## 6. CONCLUSIONS

A new method to reconstruct and identify muons with high efficiency and achieve high pion rejection in the ILC muon detector for the SiD concept has been presented. Jets physics plays an important role at the ILC,. In this paper the method has been tested on single muons and applied on bbar-b jets produced at 500 GeV . In bbar-b jets, 70% of the muons are at and above 3 GeV/c and reach the muon detector, more than 30% of them have a momentum between 3 and 7.5 GeV/c. These medium energy muons tend to lose a large proportion of their energy by dE/dx such that by the time they reach the muon detector they are 1 to 2 GeV particles. In order to identify muons in the medium energy range, a second order dE/dx corrections as well as the Kalman filter stepper were implemented in the code. The improved analysis includes now multiple scattering as well a magnetic field effects integrated using the Runge-Kutta method and the loss of energy through ionization, dE/dx. In the Kalman filter the error at the starting point has been chosen to be the angle bin size in the calorimeter ECal. Prior to the entry into ECal the particles trajectory are reconstructed by the stepper before improvements. The Kalman filter gives a realistic propagation at each step and allows hits to be collected in a narrower kinematic band. As a result of applying the Kalman Filter, the cuts around the track path may be restricted to a minimum number of angle bins, even for low momenta,. One therefore achieve a better separation of the muons from their neighbors in jets . The drop in the number of fake muons in the sample allows a better ID efficiency and an increased purity . This was previously possible only at higher momenta. This new capability is particularly important to allow for an effective separation of particles in jets.